\documentclass[a4paper,twoside,10pt]{article}
\usepackage{graphicx,publaob}
\usepackage{amssymb}

\def\papertype{Poster}
\begin{document}

\title{CONSTRAINING THE COLLECTIVE RADIO EMISSION OF LARGE SCALE ACCRETION SHOCKS}

\authors{A. \'CIPRIJANOVI\'C$^1$, T. PRODANOVI\'C$^2$ \lowercase{and} M. Z. PAVLOVI\'C$^1$}

\address{$^1$Department of Astronomy, Faculty of Mathematics, University of Belgrade, Studentski trg 16, Serbia}
\Email{aleksandra}{matf.bg.ac}{rs}
\address{$^2$Department of Physics, Faculty of Sciences, University of Novi Sad, Serbia}

\markboth{CONSTRAINING THE COLLECTIVE RADIO EMISSION OF LARGE SCALE ACCRETION SHOCKS}{}

\abstract{Accretion of gas onto already virialized structures like galaxy clusters should give rise to accretion shocks which can potentially accelerate cosmic rays. Here, we use the radio emission detected from Coma cluster and models of evolution of cosmic accretion shocks, to constrain the possible contribution of unresolved galaxy clusters to the cosmic radio background. We assume that Coma is a typical galaxy cluster and that its entire radio emission is produced by cosmic rays accelerated in accretion shocks, making our prediction an upper limit. Our models predict that at lower frequencies accretion shocks can have a potentially large contribution to the cosmic radio background, while on larger frequencies, e.g. $5\,\mathrm{GHz}$, their contribution must be lower than $\lesssim 2-35\%$, depending on the models of evolution of accretion shocks that we use.}

\section{INTRODUCTION}

The ARCADE 2 (Absolute Radiometer for Cosmology, Astrophysics and Diffuse Emission) observations of the radio sky show an excess in addition to the Cosmic Microwave Background (CMB) temperature of $T_\mathrm{CMB}=2.725\pm0.001\,\mathrm{K}$ (Fixsen et al. 2011). The existence of this excess radio emission (Cosmic Radio Background - CRB) is supported by the observations at lower frequencies (Haslam et al. 1981; Reich \& Reich 1986; Roger et al. 1999; Maeda et al. 1999). The observed excess extends from $22\,\mathrm{MHz}$ to $10\,\mathrm{GHz}$, and is well fitted by a power law $T_\mathrm{CRB} = T_\mathrm{R}\left(\frac{\nu}{310\,\mathrm{MHz}}\right)^\beta\mathrm{K}$, where $T_\mathrm{CRB}$ is the brightness temperature of the CRB, $T_\mathrm{R}=(24.1 \pm 2.1)\,\mathrm{K}$ is the normalization temperature of the CRB, $\nu$ is the frequency, and $\beta = -2.599 \pm 0.036$ is the power law index (Fixsen et al. 2011). The measured CRB is several times higher than the contribution from currently observed radio sources like galaxy clusters and the intergalactic medium, radio supernovae, radio quiet quasars, and star forming galaxies (Singal et al. 2010; Vernstrom et al. 2011). This leaves room for possible dark matter contribution (Fornengo et al. 2011; Hooper et al. 2012; Cline \& Vincent 2013) or some other unresolved radio sources.

Here, we consider cosmic-ray acceleration in large-scale accretion shocks (Miniati et al. 2000; Furlanetto \& Loeb 2004), present around galaxy clusters (Pinzke \& Pfrommer 2010). Recent detection of X-ray and gamma-ray signal around the Coma cluster could be a potential evidence for the presence of accretion shocks (Keshet et al. 2017; Keshet \& Reiss 2017). Constrains on their contribution to the gamma-ray and neutrino backgrounds are still weak, but they cannot yet be ruled out (Dobard\v zi\'c \& Prodanovi\'c 2014; 2015). Synchrotron emission from electrons accelerated in large-scale accretion shocks, should produce radio signal (Ensslin et al. 1998; Kushnir et al. 2009), but also contribute to the CRB (Keshet \& Waxman 2004).

\section{FORMALISM AND RESULTS}

We follow models from Dobard\v zi\'c \& Prodanovi\'c (2014), who have calculated the contribution of unresolved galaxy clusters to the \emph{Fermi}-LAT isotropic gamma-ray background (Ackermann et al. 2015). The observable quantity that can be compared to the CRB is the differential radio intensity  $\mathrm{d}I_\mathrm{r}/\mathrm{d}\Omega\,[\mathrm{Jy}\,\mathrm{sr}^{-1}]$ coming from all unresolved galaxy clusters:

\begin{equation}
\frac{\mathrm{d}I_{\mathrm{r}}}{\mathrm{d}\Omega} = \frac{c}{4\pi H_0 J_0(z_0)} \int_0^{z_\mathrm{vir}}
\mathrm{d}z\frac{\dot{\rho}_\mathrm{sf}(z)L_{\mathrm{r}}(\nu)}{\sqrt{\Omega_\Lambda +\Omega_\mathrm{m}(1+z)^3}}\, \nonumber\\$$
$$\times \left[ \frac{\epsilon}{\epsilon+1}+ (\epsilon+1)^{-1}\frac{\int_{z_\mathrm{vir}}^{z}\mathrm{d}z\left(\mathrm{d}t /\mathrm{d}z \right)
	\dot{\rho}_\mathrm{sf}(z)}{\int_{z_\mathrm{vir}}^{z_0}\mathrm{d}z\left(\mathrm{d}t/\mathrm{d}z\right)\dot{\rho}_\mathrm{sf}(z)} \right]\,,
\end{equation}
where $H_0$ is the present value of the Hubble parameter, $c$ speed of light, $z$ redshift and $z_\mathrm{vir}$ virialization redshift of the source, and matter and vacuum energy density parameters are $\Omega_\mathrm{m}$ and $\Omega_\Lambda$. The evolution of cosmic accretion shocks is described by the cosmic accretion rate $\dot{\rho}_\mathrm{sf}(z)\,\left[M_\odot\mathrm{yr}^{-1}\mathrm{Mpc}^{-3}\right]$ (Pavlidou \& Fields 2006). The accretion rate of a single object at redshift $z_0$ to which we normalize our models is $J_0\,[\mathrm{M_\odot\mathrm{yr}^{-1}}]$, and $\epsilon$ is the initial gas fraction of the object accreting gas. The detailed derivation of this equation and parameter values are given in Dobard\v zi\'c \& Prodanovi\'c (2014). Finally, $L_{\mathrm{r}}(\nu)\,[\mathrm{erg}\,\mathrm{s}^{-1}\mathrm{Hz}^{-1}]$ is the radio spectrum of some typical galaxy cluster. For this we use Coma cluster, since it is a well studied galaxy cluster with the observed diffuse radio emission (Large et al. 1959; Schlickeiser et al. 1987). We use fitted Coma radio spectrum by Brunetti et al. (2012), which was derived using power law in momentum $\propto p^{-2.6}$ hadronic models. Compilation of observed Coma radio data, that was used for the fitting, can be found in Pizzo (2010).

In Figure 1 we present the resulting differential radio intensity of unresolved galaxy clusters, derived using Equation (1) and integrated over the whole solid angle, $I_\mathrm{r}\,\left[\mathrm{Jy}\right]$. Dashed curve was derived using the simplest Model 1 (model depends only on the distribution of accreting objects by mass) for the evolution of accretion shocks (Pavlidou \& Fields 2006), while dotted and dash dotted curves use more realistic Models 2 and 3 (models depend on the distribution of accreting objects by mass and properties of the surrounding medium), respectively. The data points represent the CRB derived by subtracting the $T_\mathrm{CMB}$ (Fixsen et al. 2011) from the observed radio emission (Fixsen et al. 2011; Roger et al. 1999; Maeda et al. 1999; Reich \& Reich 1986; Haslam et al. 1981). The solid black line is the best fit CRB spectrum that corresponds to $T_\mathrm{CRB} \propto \nu^{-2.599}$ from Fixsen et al. (2011). Both $T_\mathrm{CRB}$ data points and the best fit spectrum were converted using $I_\mathrm{CRB}\left[\mathrm{Jy}\right] = 10^{26}\times 4\pi\times\frac{2\nu^2k T_\mathrm{CRB}}{c^2}$, where $k$ is the Boltzmann constant. Our models predict that the contribution of unresolved galaxy clusters can be high on low frequencies, although one has to keep in mind that synchrotron self-absorption will reduce the possible contribution at even lower frequencies (not included here, since these losses aren't visible in the observed Coma radio spectrum). On $5\,\mathrm{GHz}$ the contribution should be $\lesssim 2-35\%$ (upper limit range corresponds to the use of different accretion shock models). Finally, around $10\,\mathrm{GHz}$ the possible contribution sharply drops.

\begin{figure}
	\includegraphics[width=7cm]{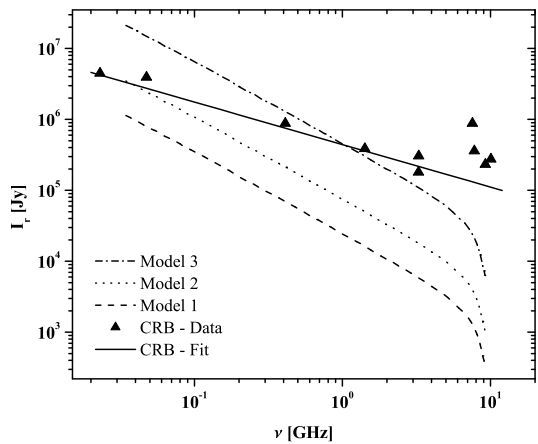}\centering
	\caption{Radio intensity of unresolved galaxy clusters derived using Model 1 for the evolution of cosmic accretion shocks (dashed line), Model 2 (dotted line), Model 3 (dash dotted line). Data points correspond to the measured CRB on various frequencies, from Fixsen et al. (2011), and the solid black line corresponds to their best fit CRB spectrum. Both data points and best fit CRB spectrum were converted to $I_\mathrm{CRB}\left[\mathrm{Jy}\right]$.}
	\label{fig:f1}
\end{figure}

\section{DISCUSSION AND CONCLUSION}

The model presented in this paper shows that large scale accretion shocks can potentially be an important contributor to the CRB. We assume that the observed radio spectrum of the Coma cluster is entirely produced by accretion shock cosmic rays, which makes our predictions an upper limit. We also assume that the Coma cluster is a typical cluster, and that clusters similar to Coma produce the bulk of the radio waves coming from large-scale accretion shocks. Our upper limits show that clusters can contribute $\lesssim 2-35\%$ at $5\,\mathrm{GHz}$, but after $10\,\mathrm{GHz}$ their contribution sharply drops.

The highest curve on Figure 1 overshoots the observed CRB at lower frequencies, which suggests that there should be a brake in the CRB spectrum at lower frequencies, which is not observed. This indicates that the Coma might not be a typical cluster, or that its entire radio emission cannot be coming from accretion shock cosmic rays. Brunetti et al. (2012) have tried to explain the Coma radio halo by synchrotron emission of secondary electrons produced via proton-proton collisions in the intra-cluster medium, or secondary electrons reaccelerated by MHD turbulence during cluster mergers. Also, not all clusters have the associated diffuse radio emission (Giovannini \& Feretti 2000; Rudnick \& Lemmerman 2009). The presence of the radio emission is often related to merging clusters and merger shocks (Fang \& Linden 2016), which are not included in our model. Contribution can also be lower if most of the radio halos have much steeper spectral indices than Coma (Liang et al. 2000).

Better understanding of accretion shocks can come from linking cluster observations at different wavelengths. Of course, one has to keep in mind that the same processes inside galaxy clusters might not be responsible for the bulk of their emitted radiation at different wavelengths. After the recent possible detection of the Coma cluster in gamma rays (Xi et al. 2017; Keshet et al. 2017) and hopefully forthcoming detections of other galaxy clusters, it will be easier to distinguish between different cosmic-ray populations inside these objects, but also better understand their possible role in the production of measured background radiation at different wavelengths.
\\
\\
\textbf{Acknowledgements} The work of A.C. and M.Z.P. is supported by the Ministry of Science of the Republic of Serbia under project number 176005, and the work of T.P. is supported in part by the Ministry of Science of the Republic of Serbia under project numbers 171002 and 176005.

\references
	
	Ackermann, M. et al.: 2015, \journal{Astrophys. J.}, \vol{799}, 86.
	
	Brunetti, G. et al.: 2012, \journal{Mon. Not. R. Astron. Soc.}, \vol{426}, 956.
	
	Cline, J.~M., Vincent, A.~C.: 2013,  \journal{J. Cosmol. Astropart. Phys.}, \vol{02}, 011.
	
	Dobard\v zi\'c, A., Prodanovi\'c, T.: 2014, \journal{Astrophys. J.}, \vol{782}, 109 [Erratum: 2014, \journal{Astrophys. J.}, \vol{787}, 95].
	
	Dobard\v zi\'c, A., Prodanovi\'c, T.: 2015, \journal{Astrophys. J.}, \vol{806}, 184.
	
	Ensslin, T.~A. et al.: 1998, \journal{Astron. Astrophys.}, \vol{332}, 395.
	
    Fang, K., Linden, T.: 2016, \journal{J. Cosmol. Astropart. Phys.}, \vol{10}, 004.

	Fixsen, D.~J. et al.: 2011, \journal{Astrophys. J.}, \vol{734}, 5.
	
	Fornengo, N. et al.: 2011, \journal{Phys. Rev. Lett.}, \vol{107}, 271302.
	
	Furlanetto, S.~R., Loeb, A.: 2004, \journal{Astrophys. J.}, \vol{611}, 642.
	
	Giovannini G., Feretti L.: 2000, \journal{New Astron.}, \vol{5}, 335.
	
	Haslam, C.~G.~T. et al.: 1981, \journal{Astron. Astrophys.}, \vol{100}, 209.
	
	Hooper, D. et al.: 2012, \journal{Phys. Rev. D}, \vol{86}, 103003.
	
	Keshet, U. et al: 2004, \journal{Astrophys. J.}, \vol{617}, 281.
	
	Keshet, U. et al.: 2017, \journal{Astrophys. J.}, \vol{845}, 24.
	
	Keshet, U., Reiss, I.: 2017, arXiv:1709.07442.
	
	Kushnir, D. et al.: 2009, \journal{J. Cosmol. Astropart. Phys.}, \vol{09}, 024.
	
	Large M. et al.: 1959, \journal{Nature}, \vol{183}, 1663L.
	
	Liang H. et al.: 2000, \journal{Astrophys. J.}, \vol{544}, 686.
	
	Maeda, K. et al.: 1999, \journal{Astron. Astrophys. Supp.}, \vol{140}, 145.
	
	Miniati, F. et al.: 2000, \journal{Astrophys. J.}, \vol{542}, 608.
	
	Pavlidou, V., Fields, B.~D.: 2006, \journal{Astrophys. J.}, \vol{642}, 734.
	
    Pinzke, A., Pfrommer, C.: 2010, \journal{Mon. Not. R. Astron. Soc.}, \vol{409}, 449.

    Pizzo, R.: 2010, \journal{Tomography of galaxy clusters through low-frequency radio polarimetry}, PhD Thesis, Groningen University.

	Reich, P., Reich, W.: 1986,  \journal{Astron. Astrophys. Supp.}, \vol{63}, 205.
	
	Roger, R.~S. et al.: 1999, \journal{Astron. Astrophys. Supp.}, \vol{137}, 7.
	
	Rudnick L., Lemmerman J.~A.: 2009, \journal{Astrophys. J.}, \vol{697}, 1341.
	
	Schlickeiser R. et al.: 1987, \journal{Astron. Astrophys.}, \vol{182}, 21.
	
	Singal, J. et al.: 2010, \journal{Mon. Not. R. Astron. Soc.}, \vol{409}, 1172.
	
	Vernstrom, T. et al.: 2011, \journal{Mon. Not. R. Astron. Soc.}, \vol{415}, 3641.
	
	Xi, S.-Q. et al.: 2017, arXiv:1709.08319.
	
\endreferences

\end{document}